\documentclass[
superscriptaddress,
nofootinbib,
 amsmath,amssymb,
 aps,
prb,
twocolumn]{revtex4-2}

\usepackage{graphicx}
\usepackage{dcolumn}
\usepackage{bm}

\usepackage{amssymb}
\usepackage{amsmath}
\usepackage{amstext}
\usepackage{graphicx,epsfig}
\usepackage{epsfig}
\usepackage{verbatim} 
\usepackage{fancyhdr}
\usepackage{fancybox}
\usepackage{color}
\usepackage{ulem,bbold}
\usepackage{enumitem}
\usepackage{caption}
\usepackage{subcaption}
\usepackage{bbm}
\usepackage{parskip}

\newcommand{\Comment}[1]{{}}
\definecolor{MyDarkBlue}{rgb}{0.15,0.15,0.45}
\usepackage[linktocpage=true]{hyperref}
\hypersetup{
colorlinks=true,
citecolor=MyDarkBlue,
linkcolor=MyDarkBlue,
urlcolor=MyDarkBlue,
}

\newcommand{\be}{\begin{equation}}
\newcommand{\ee}{\end{equation}}
\newcommand{\bea}{\begin{eqnarray}}
\newcommand{\eea}{\end{eqnarray}}

\begin{document}


\title{Applications of de Sitter Causality Conditions}

\author{Daniel McLoughlin}
 \email{dcm2183@columbia.edu}
\affiliation{Department of Physics, Columbia University,\\ New York, NY 10027, USA}
\author{Rachel A Rosen}
 \email{rarosen@cmu.edu}
\affiliation{Department of Physics, Carnegie Mellon University,\\ Pittsburgh, PA 15213, USA}

\begin{abstract}

We find explicit de Sitter shockwave solutions in arbitrary spacetime dimensions.  We use these to determine the dimensional-dependent ``stretching" of the de Sitter Penrose diagram in the presence of a shock or black hole.  This stretching sets the scale at which superluminalities in de Sitter can be considered resolvable.  We then consider an $RFF$ coupling for a spin-1 particle and a Gauss-Bonnet coupling for a spin-2 particle and analyze these in terms of our proposed de Sitter causality criteria.  We briefly discuss the connections between our results and corresponding results in anti de Sitter spacetime.


\end{abstract}

\maketitle


\section{Introduction}
\label{sec:intro}
\vspace{-0.25cm}
Causality conditions place powerful constraints on low energy effective field theories (see, e.g., \cite{Adams:2006sv,Camanho:2014apa,Hartman:2015lfa}, as well as \cite{deRham:2022hpx} and references therein).  They help us understand the space of allowed field theories, to determine which theories are UV completable and to understand the spectrum of low energy states.  While significant advances have been made recently in constraining theories via causality conditions in flat spacetime and in anti de Sitter spacetime (via constraints on the space of consistent CFTs from the conformal bootstrap), much less progress has been made in constraining theories in de Sitter spacetime.  There are many reasons for this, among them the lack of a de Sitter S-matrix (although, see, e.g., \cite{Bousso:2004tv,Spradlin:2001nb,Marolf:2012kh,Melville:2023kgd,Melville:2024ove}) as well as the difficulty in defining a time delay or advance on the space-like boundary of de Sitter spacetime.

In previous work \cite{Bittermann:2022hhy}, we proposed a notion of causality in de Sitter spacetime by using spatial shifts of the end points of null geodesics measured on the boundary of de Sitter.  Our motivation came from \cite{Gao:2000ga} where it was proven that, in a spacetime with a timelike conformal boundary satisfying strong causality, the ``fastest null geodesic" connecting two points on the boundary must run entirely along the boundary and not through the bulk. In other words, the causal structure on the boundary sets the maximum speed for propagation in the bulk.  While this proof obviously does not hold for asymptotically de Sitter spacetimes, we argued that the closest analogous condition would be to require the fastest null geodesics in de Sitter be those at maximum impact parameter from a gravitational source, i.e., at $b=1/H$.  This condition can be imposed unambiguously by considering the spatial positions of the endpoints of complete null geodesics on the past and future boundaries of de Sitter.

When computing the Shapiro time delay in de Sitter spacetime due to a black hole or a shockwave, one naively finds a small time advance at large impact parameter.  However, as we argued in \cite{Bittermann:2022hhy}, this is because the result is capturing two competing effects: the usual slowing of a light ray due to a gravitational source and the ``stretching" of the Penrose diagram -- a phenomena specific to asymptotically de Sitter spacetimes \cite{Gao:2000ga}.  By using the null geodesic at largest impact parameter to set the causal structure, one can subtract off the effect of the stretching.  Small time ``advances" at large impact parameter are then no longer resolvable while large time advances at small impact parameter remain pathological.

 In this paper we extend the considerations of our previous work to arbitrary spacetime dimensions.  In particular, we calculate the $D$-dimensional de Sitter shockwave solutions as well as the boundary spatial shift experienced by a null geodesic traversing this shockwave.  This allows us to, in effect, compute the dimension-dependent stretching of de Sitter spacetime due to a shockwave or, equivalently, a black hole.\footnote{The stretch is the same in both cases as the two configurations are related by an ultrarelativistic boost \cite{Aichelburg:1970dh, Hotta:1992qy}.}  We next consider non-minimal couplings of spin-1 and spin-2 particles, the $RFF$ and Gauss-Bonnet terms respectively, and show how causality violation can be seen in de Sitter in close analogue to the flat spacetime case discussed in \cite{Camanho:2014apa}.  We briefly discuss the relationship between our results and those in anti de Sitter spacetime and end with some discussion of future directions.

\section{de Sitter Shock Wave Solutions}
\label{sec:shock}
\vspace{-0.25cm}

We begin by deriving de Sitter shockwave solutions in arbitrary spacetime dimensions.  Shockwave solutions in de Sitter spacetime were first found in \cite{Hotta:1992qy} in four spacetime dimensions. Our results are consistent with those solutions in the $D=4$ case and with an appropriate change of coordinates.

Our formalism closely follows that used in \cite{Bittermann:2022hhy}. In particular, to study the shock wave, we introduce lightcone coordinates $u,v$ in the $t-z$ plane of the de Sitter static coordinates:
\begin{subequations}
\label{ttou}
\begin{align}
&t= \frac{1}{2 H} \log \left[ \frac{(1+Hu )(1+Hv)}{(1-Hu)(1-Hv)}   \right] \, ,\\
&z = \sqrt{1-H^2b^2}\,\frac{v-u}{1-H^2u v} \, .
\end{align}
\end{subequations}
In these coordinates, the pure de Sitter metric becomes
\be
\label{pen}
ds_{{\rm dS}}^{2}=-4\frac{1-H^{2}b^{2}}{\left(1-H^{2}uv\right)^{2}}\,du\,dv+\frac{db^{2}}{1-H^{2}b^{2}}+b^{2}d\Omega_{D-3}^{2}\, ,
\ee
where $d\Omega_{D-3}^{2}$ is the metric on the $\left(D-3\right)$-sphere and $b$ is the impact parameter $r^2 = b^2+z^2$. 

For the metric in the presence of a shock wave we adopt a Kerr-Schild ansatz:
\be
\label{dS}
ds^2 = ds_{\rm dS}^2 + \Delta ds^2 \, ,
\ee
with the metric perturbation given by
\be
\label{lc}
\Delta ds^2 =  F[x] (k_\mu dx^\mu)^2\, ,
\ee
where $k_\mu= (0,-1,\vec{0})$.   With this ansatz, the general equations of motion can be written as
\be
\Box h^\mu_{~\nu}-2 H^2 h^\mu_{~\nu} = -16 \pi G\, T^\mu_{~\nu} \, ,
\ee
where $h^\mu_{~\nu}\equiv F[v,\vec{x}]k^\mu k_\nu$.  The source is taken to be that of a point particle moving at the speed of light with momentum $p_u$:
\be
T_{\mu\nu}=p_u\,\delta(v)\delta(b) k_\mu k_\nu \, .
\ee
\begin{widetext}
Taking $F[v,\vec{x}] \rightarrow 2Gp_u\,f(b)\,\delta(v) $, the equations of motion become
\be
\label{eom}
\left(1-H^2b^2\right)f''(b)+\frac{1}{b}\left((D-3)(1-H^2b^2)+H^2b^2 \right)f'(b)+2H^2\left((D-3)+\frac{1}{1-H^2 b^2} \right)f(b)= -8  \pi \,\delta(\vec{x}) \, .
\ee
Fixing appropriate boundary conditions, the solution for the shock profile is given by
\be
\label{Fb}
f(b)=-\frac{4 H^{D-4}\, \Gamma \left[\frac{1}{2}(D-2) \right] }{\pi^{\frac{D-4}{2}}}  \sqrt{1-H^{2}b^{2}} \,{}_2F_1\left(-\frac{1}{2},\frac{1}{2}\left(D-2\right);\frac{1}{2};1-H^{2}b^{2}\right)  \, .
\ee
In these coordinates, the flat space limit $H \rightarrow 0$ can be readily obtained and this solution agrees with the those found previously in the literature, see, e.g., \cite{Camanho:2014apa}.\footnote{There is an overall factor of 4 discrepancy due to the different normalization of $u$ and $v$ as defined in \eqref{ttou} compared to \cite{Camanho:2014apa}.}  In particular, for the $D=4$ case which is known to be logarithmically divergent in the IR in flat spacetime, the de Sitter scale $H$ serves as an IR cutoff: $f(b) \rightarrow 4 \left( \log \left[\frac{2}{b H}\right] -1 \right)$.
\end{widetext}

We now consider the shift in the $u$-coordinate experienced by a null geodesic traveling in the $v$ direction at impact parameter $b$ and crossing the shock wave at $v=0$. Setting $ds^2 =0$ for the metric \eqref{dS} and solving for $du$, we see that the shift is given by
\be
\label{delu}
\Delta u =2Gp_u\frac{f(b)}{4(1-H^2 b^2)}\,  \, .
\ee
We next rotate our system so that the null geodesic passes though the origin and the shock wave travels past at impact parameter $b$.  As is derived in our previous work \cite{Bittermann:2022hhy}, the result is an overall factor of $\sqrt{1-H^2b^2}$ multiplying expression \eqref{delu} so that
\be
\label{delu2}
\Delta u \rightarrow 2Gp_u\frac{f(b)}{4\sqrt{1-H^2 b^2}}\,  \, .
\ee
This quantity $\Delta u$ can be interpreted as a spatial shift on the future boundary of de Sitter spacetime.  To make this explicit, we transform this shift in $u$ to a shift of the spatial conformal coordinate $\chi$ on the boundary of de Sitter using the conformal coordinates
\be
\label{conf}
ds^2 =\frac{1}{\cos^2HT}  \left(-dT^2 +\frac{1}{H^2}d\chi^2+\frac{1}{H^2}\sin^2\chi \, d\Omega^2\right) \, ,
\ee
where
\be
-\frac{\pi}{2} \leq HT \leq \frac{\pi}{2} \, , ~~~~ {\rm and} ~~~~ 0 \leq \chi \leq \pi \, .
\ee
For small $\Delta u$, the resulting shift in $\chi$ is given by
\be
\Delta \chi \simeq - H\Delta u \, .
\ee
The expression for the shift in the presence of the shock is thus
\be
\label{dchi}
  \Delta \chi = 
  (H r_s)^{D-3}\,
  \frac{\sqrt{\pi}\, \Gamma\left[\frac{D}{2}\right]}{2\,\Gamma\left[\frac{D-1}{2}\right]}\, 
{}_2F_1\left(-\tfrac{1}{2},\tfrac{D-2}{2};\tfrac{1}{2};1-H^{2}b^{2}\right)   .  
\ee
where, in analogy with the $D-$dimensional Schwarzschild radius, we have defined
\be
r_s^{D-3} = \frac{16 \pi G \,p_u}{D-2} \frac{\Gamma\left[\frac{D-1}{2}\right]}{2 \pi^{
\frac{D-1}{2}}} \, .
\ee

The shift in $\chi$ at the future boundary of de Sitter is not sign definite.  At small impact parameter, the shift is negative, implying that the lightcone of this geodesic lies inside the original lightcone given by a null geodesic in empty de Sitter.  As the impact parameter increases, the the shift becomes positive, naively implying that this geodesic lies outside the original lightcone of empty de Sitter.  As was argued in \cite{Bittermann:2022hhy}, this apparent superluminality is because two competing effects are reflected in this value: the usual Shapiro time delay which would correspond to negative $\Delta \chi$ and the ``stretching" of the de Sitter Penrose diagram in the presence of bulk matter \cite{Gao:2000ga} which pushes $\Delta \chi$ towards the positive direction.  In order to consider causality, we can subtract off the effect of the stretching.

To do so, we note that in de Sitter spacetime there is a maximum possible impact parameter $b=1/H$.  Beyond this, the shockwave enters
the southern hemisphere of the de Sitter sphere.  Accordingly, there is a maximum possible positive shift on the boundary given by
\be
\label{maxShift}
\Delta\chi_{max} = (H r_s)^{D-3}\,\frac{\sqrt{\pi}}{2}\frac{\Gamma\left[\frac{D}{2}\right]}{\Gamma\left[\frac{D-1}{2}\right]} \, .
\ee
Let us use the geodesic at maximum impact parameter in the presence of the shock to set the causal structure.  I.e., in a Penrose diagram, this geodesic is fixed to travel along a $45^{\circ}$ angle.  Thus, to account for the positive shift in $\Delta \chi$ seen in \eqref{maxShift}, the de Sitter Penrose diagram must stretch in the time-like direction $T$ by precisely this amount.  Because of this, we can think of superluminality in de Sitter spacetime as only being resolvable up to the amount given in \eqref{maxShift}.

\section{Non-Minimal Coupling: Spin-1}
\label{sec:spin1}
\vspace{-0.25cm}

We have argued that a small amount of positive spatial shift in the conformal coordinate $\chi$ at the boundary of de Sitter for large impact parameters $b$ is not necessarily indicative of a pathology.  We next consider theories that exhibit large positive shifts of $\chi$ at small impact parameter and argue that this behavior is what captures superluminality.

As a first example we consider an $RFF$ correction to pure Maxwell theory on the de Sitter shockwave background, with an action given by
\begin{equation}
\label{RFF}
S=\int\mathrm{d}^{D}x\sqrt{-g}\left(-\frac{1}{4}F_{\mu\nu}F^{\mu\nu}+\frac{\alpha}{4\,\Lambda^2}\hat{R}_{\alpha\beta}^{\;\;\;\;\mu\nu}F^{\alpha\beta}F_{\mu\nu}\right).
\end{equation}
The coupling here is to the subtracted Riemann tensor defined via
\begin{equation}
\hat{R}_{\alpha\beta}^{\;\;\;\;\mu\nu}\equiv R_{\alpha\beta}^{\;\;\;\;\mu\nu}-\bar{R}_{\alpha\beta}^{\;\;\;\;\mu\nu},
\end{equation}
where $\bar{R}_{\alpha\beta}^{\;\;\;\;\;\mu\nu}$ is the Riemann
tensor of pure de Sitter spacetime. On-shell this becomes the Weyl
tensor \cite{Horowitz:1999gf}.  The presence of time advances for this model in flat spacetime was analyzed in \cite{Camanho:2014apa}. Later, in de Sitter spacetime and in $D=4$, the analogous result was analyzed in \cite{Bittermann:2022hhy}.  Here, we generalize the latter result to arbitrary spacetime dimensions.

Our goal is to calculate the correction to the shift in the lightcone coordinate $u$ of a spin-1 particle traversing a shockwave due to the non-minimal coupling.  The equations of motion that follow from \eqref{RFF} are given by
\begin{equation}
  \nabla^{\mu}(F_{\mu \nu}-\frac{\alpha}{\Lambda^2} \hat{R}_{\mu\nu}^{~~\alpha\beta}F_{\alpha\beta}) = 0 \, .
\end{equation}

\begin{widetext}

We solve the equations of motion for the physical polarizations of the spin-1 field in the de Sitter shockwave background.  After solving, we again rotate our system so that we are describing a light ray that passes through the origin of our system and a shockwave at impact parameter $b$ (rather than the other way around).  We isolate the polarizations along the impact parameter, $\vec{\epsilon} = \vec{b}/b$, and orthogonal to the impact parameter, $\vec{\epsilon}\cdot\vec{b}=0$.  We find the following shifts due to the presence of the shock:
\begin{equation}
\Delta u=2Gp_u \left[\frac{f\left(b\right)}{4\sqrt{1-H^{2}b^{2}}}+\frac{\alpha}{4\Lambda^2}\sqrt{1-H^{2}b^{2}}\times\begin{cases}
\frac{2H^{2}f\left(b\right)}{\left(1-H^{2}b^{2}\right)^{2}}+\frac{H^{2}bf^{\prime}\left(b\right)}{\left(1-H^{2}b^{2}\right)}+f^{\prime\prime}\left(b\right) \, ,& \mathrm{for}\;\vec{\epsilon} \parallel \vec{b}\\ 
\frac{2H^{2}f\left(b\right)}{\left(1-H^{2}b^{2}\right)}+\frac{f^{\prime}\left(b\right)}{b} \, ,& \mathrm{for}\;\vec{\epsilon} \perp \vec{b}
\end{cases} \right] \, .
\end{equation}

The correction to the shift due to the non-minimal coupling simplifies greatly using the explicit form of $f(b)$ found above in equation \eqref{Fb} as well as using the properties of the hypergeometric functions.  With some massaging, we can write the $D$-dimensional shift as
\begin{equation}
\label{Du}
\Delta u=2Gp_u\left[ \frac{f\left(b\right)}{4\sqrt{1-H^{2}b^{2}}}+\frac{\alpha}{\Lambda^2} \,\frac{ \Gamma \left[\frac{1}{2}(D-2) \right]}{\pi^{\frac{D-4}{2}}b^{D-2}} \times \begin{cases}
D-3, & \mathrm{for}\;\vec{\epsilon} \parallel \vec{b}\\
-1, & \mathrm{for}\;\vec{\epsilon} \perp \vec{b}
\end{cases} \right] \, .
\end{equation}
 Notably, while the shift of a null geodesic in the de Sitter shockwave background depends explicitly on the de Sitter scale $H$, the correction term due to the $RFF$ coupling is manifestly independent of $H$.  In other words, the correction in de Sitter is identical to the correction in flat spacetime.  This is follows from having effectively coupled the spin-1 field to the Weyl tensor.  

Before analyzing further, let us consider the flat spacetime limit of the above expression and expand in small $H$:
\begin{equation}
\Delta u\simeq 2Gp_u \frac{\Gamma\left[\frac{D-4}{2}\right]}{2 \pi^{\frac{D-4}{2}}} \frac{1}{b^{D-4}}\left[ 1+\frac{D-3}{D-6}\, b^2 H^2+{\cal O}(b^4 H^4)+\alpha \,\frac{D-4}{\Lambda^2 b^2} \times \begin{cases}
D-3, & \mathrm{for}\;\vec{\epsilon} \parallel \vec{b}\\
-1, & \mathrm{for}\;\vec{\epsilon} \perp \vec{b}
\end{cases} \right] \, .
\end{equation}
Here the $D=4$ and $D=6$ cases need to be derived specially from equation \eqref{Du}.

\end{widetext}

In order to associate physical meaning to this shift we relate it to a shift of the spatial coordinate $\chi$ on the future boundary of our de Sitter spacetime: $\Delta \chi \simeq - H\Delta u$.  We see that, as in the flat spacetime case, because the two polarizations of the photon receive corrections with opposite signs, then at sufficiently small impact parameter $b \sim 1/\Lambda$, the shifts $\Delta \chi$ corresponding to each polarization will have opposite signs, independent of the sign of $\alpha$.  Moreover, one can make the positive shift in $\Delta \chi$ for one polarization arbrtrarily large at small enough impact parameter so that its path lies outside the light cone of the fiducial ``fastest'' geodesic at maximum impact parameter.  In other words, the positive $\Delta \chi$ at small impact parameter can be made larger than the small positive shift \eqref{maxShift} at the maximum impact parameter.  The $RFF$ coupling can thus be seen to violate our de Sitter causality criteria.

\section{Non-Minimal Coupling: Spin-2}
\label{sec:spin2}
\vspace{-0.25cm}

We next consider a non-minimal coupling for a spin-2 particle in de Sitter spacetime given by the Gauss-Bonnet term. Superluminalities due to the presence of such a term have been studied previously in the literature, in both flat and AdS spacetimes. Here we follow the analysis of \cite{Camanho:2014apa}. The action is given by
\begin{equation}
S=\int\mathrm{d}^{D}x\sqrt{-g}\left(R+\lambda\left[R_{\mu\nu\rho\sigma}R^{\mu\nu\rho\sigma}-4 R_{\mu\nu}R^{\mu\nu}+R^{2}\right]\right).
\end{equation}
Our interest is the correction to the shift in the lightcone coordinate $u$ of a spin-2 perturbation traversing the de Sitter shockwave background.  

Following \cite{Camanho:2014apa}, we consider only those contributions to the lightcone shift in the transverse equations of motion and we ignore terms that shift all polarizations equally, leading to the equations of motion
\begin{equation}
\delta R_{\mu\nu}+\lambda\hat{R}_{(\mu}^{\quad\rho\alpha\beta}\delta R_{\nu)\rho\alpha\beta}=0 \, ,
\end{equation}
where again $\hat{R}$ is the subtracted Riemann tensor.  

We solve the equations of motion for the shift, multiply by the rotation factor and substitute in the explicit form for the de Sitter shockwave $f(b)$ as we did in the spin-1 case.   The shifts for the relevant polarizations are then found to be identical to the spin-1 case, given by
\begin{equation}
\Delta u=2Gp_u\left[\frac{f\left(b\right)}{4\sqrt{1-H^{2}b^{2}}}+\lambda\,\frac{\Gamma \left[\frac{1}{2}(D-2) \right]}{\pi^{\frac{D-4}{2}}b^{D-2}}\times \begin{cases}
D-3 \!\!\!\!\!\! \\
-1
\end{cases}\right] .
\end{equation}
The relevant polarizations of the spin-2 perturbations thus exhibit the same superluminality at small impact parameter as the spin-1 particle, by the standard of our de Sitter causality criteria.  

The correction terms due to the Gauss-Bonnet coupling are also manifestly independent of the de Sitter scale $H$ and are identical to the flat spacetime correction terms.  In contrast, higher derivative correction terms, such as $\nabla_{(\mu}\nabla_{\nu)}\hat{R}_{\alpha\beta}^{\;\;\;\;\gamma\delta}\delta R^{\alpha\beta}_{\;\;\;\;\gamma\delta}$ considered in \cite{Camanho:2014apa} that result from a $\left(\mathrm{Riemann}\right)^{3}$ term in the action, generally yield $H$-dependent corrections.

\section{Connection to AdS}
\label{sec:AdS}
\vspace{-0.25cm}

Let us comment on the relationship between our results and those found in the literature for anti de Sitter spacetimes, for example in \cite{Camanho:2014apa} and \cite{Gubser:2008pc}.  The differential equation for the de Sitter shockwave \eqref{eom} can be analytically continued via the usual identification $H\rightarrow\frac{i}{L}$ in order to obtain the correct differential equation for the AdS shock.  In addition, the coordinates of this paper are related to those of \cite{Camanho:2014apa} via the identification $\frac{b^2}{L^2} = \frac{4 \rho^2}{\left(1 -\rho^2\right)^2}$ and to those of \cite{Gubser:2008pc} via the identification $\frac{b^2}{L^2} = 4q(1+q)$.  

However, the de Sitter and AdS shockwave solutions are not related by a simple analytic continuation as they require different boundary conditions.  Namely, in AdS one imposes the boundary condition that $f\left(b\right)\rightarrow0$ as $b\rightarrow\infty$.  In de Sitter, $b$ is not well-defined at infinity and one must instead enforce regularity of the shock at the de Sitter horizon.  Thus naively analytically continuing the de Sitter shockwave solution \eqref{Fb} to AdS results in a complex and spurious term and not the result of \cite{Camanho:2014apa} and \cite{Gubser:2008pc}.

While the shockwave solutions themselves for de Sitter and AdS cannot be related via direct analytic continuation, the corrections to the time delays that are produced by the particular non-minimal couplings considered here are actually identical.   In other words, in an AdS background, the $RFF$ and Gauss-Bonnet corrections to the shifts are also independent of the background curvature scale $L$ when evaluated on the AdS shockwave solution.  This can be verified in \cite{Camanho:2014apa} by using the explicit form of the AdS shock.

We further note that, as was discussed in \cite{Bittermann:2022hhy}, in AdS spacetime the positivity of the shift of the time coordinate $\Delta t$ on the AdS boundary can be directly linked to the positivity of the average null energy taken along the geodesic traversing the shock.  In de Sitter spacetime, this relation is not sign definite and the average null energy can be positive while the spatial shift on the de Sitter boundary $\Delta \chi$ is positive or negative.

\section{Discussion}
\label{sec:Disc}
\vspace{-0.25cm}

This work on de Sitter causality is undoubtedly still quite preliminary.  We have used a proposed notion of causality in de Sitter spacetime to analyze the now canonical cases of non-minimal coupling for spin-1 and spin-2 particles in arbitrary spacetime dimension.  It would be interesting to use the considerations of this paper to put constraints on higher derivative terms in UV consistent gravitational theories in de Sitter spacetime, as was done in \cite{Caron-Huot:2021rmr} in flat spacetime.  Indeed, there the logarithmic IR divergence present in $D=4$ in the eikonal limit (which is what is captured by phase shift traversing a shock, see, e.g., \cite{Kabat:1992tb}) makes it difficult to place a rigorous positivity bound on the coefficient of a $(\partial \phi)^4$ term.  In de Sitter spacetime this IR divergence is regulated by the de Sitter scale.  Another direction of interest would be to see how the ``stringy" equivalence principle proposed in \cite{Kologlu:2019bco} plays out in de Sitter spacetime.

\begin{acknowledgments}
We wish to thank Brando Bellazzini, Riccardo Gonzo and Scott Melville for interesting and useful conversations. We also wish to thank KITP where this work was completed: this research was supported in part by grant NSF PHY-2309135 to the Kavli Institute for Theoretical Physics (KITP).  R.A.R. is supported by the US Department of Energy grant DE-SC0010118.
\end{acknowledgments}

\bibliography{dSCaus}

\begin{thebibliography}{18}%
\makeatletter
\providecommand \@ifxundefined [1]{%
 \@ifx{#1\undefined}
}%
\providecommand \@ifnum [1]{%
 \ifnum #1\expandafter \@firstoftwo
 \else \expandafter \@secondoftwo
 \fi
}%
\providecommand \@ifx [1]{%
 \ifx #1\expandafter \@firstoftwo
 \else \expandafter \@secondoftwo
 \fi
}%
\providecommand \natexlab [1]{#1}%
\providecommand \enquote  [1]{``#1''}%
\providecommand \bibnamefont  [1]{#1}%
\providecommand \bibfnamefont [1]{#1}%
\providecommand \citenamefont [1]{#1}%
\providecommand \href@noop [0]{\@secondoftwo}%
\providecommand \href [0]{\begingroup \@sanitize@url \@href}%
\providecommand \@href[1]{\@@startlink{#1}\@@href}%
\providecommand \@@href[1]{\endgroup#1\@@endlink}%
\providecommand \@sanitize@url [0]{\catcode `\\12\catcode `\$12\catcode
  `\&12\catcode `\#12\catcode `\^12\catcode `\_12\catcode `\%12\relax}%
\providecommand \@@startlink[1]{}%
\providecommand \@@endlink[0]{}%
\providecommand \url  [0]{\begingroup\@sanitize@url \@url }%
\providecommand \@url [1]{\endgroup\@href {#1}{\urlprefix }}%
\providecommand \urlprefix  [0]{URL }%
\providecommand \Eprint [0]{\href }%
\providecommand \doibase [0]{https://doi.org/}%
\providecommand \selectlanguage [0]{\@gobble}%
\providecommand \bibinfo  [0]{\@secondoftwo}%
\providecommand \bibfield  [0]{\@secondoftwo}%
\providecommand \translation [1]{[#1]}%
\providecommand \BibitemOpen [0]{}%
\providecommand \bibitemStop [0]{}%
\providecommand \bibitemNoStop [0]{.\EOS\space}%
\providecommand \EOS [0]{\spacefactor3000\relax}%
\providecommand \BibitemShut  [1]{\csname bibitem#1\endcsname}%
\let\auto@bib@innerbib\@empty
\bibitem [{\citenamefont {Adams}\ \emph {et~al.}(2006)\citenamefont {Adams},
  \citenamefont {Arkani-Hamed}, \citenamefont {Dubovsky}, \citenamefont
  {Nicolis},\ and\ \citenamefont {Rattazzi}}]{Adams:2006sv}%
  \BibitemOpen
  \bibfield  {author} {\bibinfo {author} {\bibfnamefont {A.}~\bibnamefont
  {Adams}}, \bibinfo {author} {\bibfnamefont {N.}~\bibnamefont {Arkani-Hamed}},
  \bibinfo {author} {\bibfnamefont {S.}~\bibnamefont {Dubovsky}}, \bibinfo
  {author} {\bibfnamefont {A.}~\bibnamefont {Nicolis}},\ and\ \bibinfo {author}
  {\bibfnamefont {R.}~\bibnamefont {Rattazzi}},\ }\bibfield  {title} {\bibinfo
  {title} {{Causality, analyticity and an IR obstruction to UV completion}},\
  }\href {https://doi.org/10.1088/1126-6708/2006/10/014} {\bibfield  {journal}
  {\bibinfo  {journal} {JHEP}\ }\textbf {\bibinfo {volume} {10}},\ \bibinfo
  {pages} {014}},\ \Eprint {https://arxiv.org/abs/hep-th/0602178}
  {arXiv:hep-th/0602178} \BibitemShut {NoStop}%
\bibitem [{\citenamefont {Camanho}\ \emph {et~al.}(2016)\citenamefont
  {Camanho}, \citenamefont {Edelstein}, \citenamefont {Maldacena},\ and\
  \citenamefont {Zhiboedov}}]{Camanho:2014apa}%
  \BibitemOpen
  \bibfield  {author} {\bibinfo {author} {\bibfnamefont {X.~O.}\ \bibnamefont
  {Camanho}}, \bibinfo {author} {\bibfnamefont {J.~D.}\ \bibnamefont
  {Edelstein}}, \bibinfo {author} {\bibfnamefont {J.}~\bibnamefont
  {Maldacena}},\ and\ \bibinfo {author} {\bibfnamefont {A.}~\bibnamefont
  {Zhiboedov}},\ }\bibfield  {title} {\bibinfo {title} {{Causality Constraints
  on Corrections to the Graviton Three-Point Coupling}},\ }\href
  {https://doi.org/10.1007/JHEP02(2016)020} {\bibfield  {journal} {\bibinfo
  {journal} {JHEP}\ }\textbf {\bibinfo {volume} {02}},\ \bibinfo {pages}
  {020}},\ \Eprint {https://arxiv.org/abs/1407.5597} {arXiv:1407.5597 [hep-th]}
  \BibitemShut {NoStop}%
\bibitem [{\citenamefont {Hartman}\ \emph {et~al.}(2016)\citenamefont
  {Hartman}, \citenamefont {Jain},\ and\ \citenamefont
  {Kundu}}]{Hartman:2015lfa}%
  \BibitemOpen
  \bibfield  {author} {\bibinfo {author} {\bibfnamefont {T.}~\bibnamefont
  {Hartman}}, \bibinfo {author} {\bibfnamefont {S.}~\bibnamefont {Jain}},\ and\
  \bibinfo {author} {\bibfnamefont {S.}~\bibnamefont {Kundu}},\ }\bibfield
  {title} {\bibinfo {title} {{Causality Constraints in Conformal Field
  Theory}},\ }\href {https://doi.org/10.1007/JHEP05(2016)099} {\bibfield
  {journal} {\bibinfo  {journal} {JHEP}\ }\textbf {\bibinfo {volume} {05}},\
  \bibinfo {pages} {099}},\ \Eprint {https://arxiv.org/abs/1509.00014}
  {arXiv:1509.00014 [hep-th]} \BibitemShut {NoStop}%
\bibitem [{\citenamefont {de~Rham}\ \emph {et~al.}(2022)\citenamefont
  {de~Rham}, \citenamefont {Kundu}, \citenamefont {Reece}, \citenamefont
  {Tolley},\ and\ \citenamefont {Zhou}}]{deRham:2022hpx}%
  \BibitemOpen
  \bibfield  {author} {\bibinfo {author} {\bibfnamefont {C.}~\bibnamefont
  {de~Rham}}, \bibinfo {author} {\bibfnamefont {S.}~\bibnamefont {Kundu}},
  \bibinfo {author} {\bibfnamefont {M.}~\bibnamefont {Reece}}, \bibinfo
  {author} {\bibfnamefont {A.~J.}\ \bibnamefont {Tolley}},\ and\ \bibinfo
  {author} {\bibfnamefont {S.-Y.}\ \bibnamefont {Zhou}},\ }\bibfield  {title}
  {\bibinfo {title} {{Snowmass White Paper: UV Constraints on IR Physics}},\
  }in\ \href@noop {} {\emph {\bibinfo {booktitle} {{Snowmass 2021}}}}\
  (\bibinfo {year} {2022})\ \Eprint {https://arxiv.org/abs/2203.06805}
  {arXiv:2203.06805 [hep-th]} \BibitemShut {NoStop}%
\bibitem [{\citenamefont {Bousso}(2005)}]{Bousso:2004tv}%
  \BibitemOpen
  \bibfield  {author} {\bibinfo {author} {\bibfnamefont {R.}~\bibnamefont
  {Bousso}},\ }\bibfield  {title} {\bibinfo {title} {{Cosmology and the
  S-matrix}},\ }\href {https://doi.org/10.1103/PhysRevD.71.064024} {\bibfield
  {journal} {\bibinfo  {journal} {Phys. Rev. D}\ }\textbf {\bibinfo {volume}
  {71}},\ \bibinfo {pages} {064024} (\bibinfo {year} {2005})},\ \Eprint
  {https://arxiv.org/abs/hep-th/0412197} {arXiv:hep-th/0412197} \BibitemShut
  {NoStop}%
\bibitem [{\citenamefont {Spradlin}\ and\ \citenamefont
  {Volovich}(2002)}]{Spradlin:2001nb}%
  \BibitemOpen
  \bibfield  {author} {\bibinfo {author} {\bibfnamefont {M.}~\bibnamefont
  {Spradlin}}\ and\ \bibinfo {author} {\bibfnamefont {A.}~\bibnamefont
  {Volovich}},\ }\bibfield  {title} {\bibinfo {title} {{Vacuum states and the S
  matrix in dS / CFT}},\ }\href {https://doi.org/10.1103/PhysRevD.65.104037}
  {\bibfield  {journal} {\bibinfo  {journal} {Phys. Rev. D}\ }\textbf {\bibinfo
  {volume} {65}},\ \bibinfo {pages} {104037} (\bibinfo {year} {2002})},\
  \Eprint {https://arxiv.org/abs/hep-th/0112223} {arXiv:hep-th/0112223}
  \BibitemShut {NoStop}%
\bibitem [{\citenamefont {Marolf}\ \emph {et~al.}(2013)\citenamefont {Marolf},
  \citenamefont {Morrison},\ and\ \citenamefont {Srednicki}}]{Marolf:2012kh}%
  \BibitemOpen
  \bibfield  {author} {\bibinfo {author} {\bibfnamefont {D.}~\bibnamefont
  {Marolf}}, \bibinfo {author} {\bibfnamefont {I.~A.}\ \bibnamefont
  {Morrison}},\ and\ \bibinfo {author} {\bibfnamefont {M.}~\bibnamefont
  {Srednicki}},\ }\bibfield  {title} {\bibinfo {title} {{Perturbative S-matrix
  for massive scalar fields in global de Sitter space}},\ }\href
  {https://doi.org/10.1088/0264-9381/30/15/155023} {\bibfield  {journal}
  {\bibinfo  {journal} {Class. Quant. Grav.}\ }\textbf {\bibinfo {volume}
  {30}},\ \bibinfo {pages} {155023} (\bibinfo {year} {2013})},\ \Eprint
  {https://arxiv.org/abs/1209.6039} {arXiv:1209.6039 [hep-th]} \BibitemShut
  {NoStop}%
\bibitem [{\citenamefont {Melville}\ and\ \citenamefont
  {Pimentel}(2024{\natexlab{a}})}]{Melville:2023kgd}%
  \BibitemOpen
  \bibfield  {author} {\bibinfo {author} {\bibfnamefont {S.}~\bibnamefont
  {Melville}}\ and\ \bibinfo {author} {\bibfnamefont {G.~L.}\ \bibnamefont
  {Pimentel}},\ }\bibfield  {title} {\bibinfo {title} {{de Sitter S matrix for
  the masses}},\ }\href {https://doi.org/10.1103/PhysRevD.110.103530}
  {\bibfield  {journal} {\bibinfo  {journal} {Phys. Rev. D}\ }\textbf {\bibinfo
  {volume} {110}},\ \bibinfo {pages} {103530} (\bibinfo {year}
  {2024}{\natexlab{a}})},\ \Eprint {https://arxiv.org/abs/2309.07092}
  {arXiv:2309.07092 [hep-th]} \BibitemShut {NoStop}%
\bibitem [{\citenamefont {Melville}\ and\ \citenamefont
  {Pimentel}(2024{\natexlab{b}})}]{Melville:2024ove}%
  \BibitemOpen
  \bibfield  {author} {\bibinfo {author} {\bibfnamefont {S.}~\bibnamefont
  {Melville}}\ and\ \bibinfo {author} {\bibfnamefont {G.~L.}\ \bibnamefont
  {Pimentel}},\ }\bibfield  {title} {\bibinfo {title} {{A de Sitter S-matrix
  from amputated cosmological correlators}},\ }\href
  {https://doi.org/10.1007/JHEP08(2024)211} {\bibfield  {journal} {\bibinfo
  {journal} {JHEP}\ }\textbf {\bibinfo {volume} {08}},\ \bibinfo {pages}
  {211}},\ \Eprint {https://arxiv.org/abs/2404.05712} {arXiv:2404.05712
  [hep-th]} \BibitemShut {NoStop}%
\bibitem [{\citenamefont {Bittermann}\ \emph {et~al.}(2023)\citenamefont
  {Bittermann}, \citenamefont {McLoughlin},\ and\ \citenamefont
  {Rosen}}]{Bittermann:2022hhy}%
  \BibitemOpen
  \bibfield  {author} {\bibinfo {author} {\bibfnamefont {N.}~\bibnamefont
  {Bittermann}}, \bibinfo {author} {\bibfnamefont {D.}~\bibnamefont
  {McLoughlin}},\ and\ \bibinfo {author} {\bibfnamefont {R.~A.}\ \bibnamefont
  {Rosen}},\ }\bibfield  {title} {\bibinfo {title} {{On causality conditions in
  de Sitter spacetime}},\ }\href {https://doi.org/10.1088/1361-6382/accc05}
  {\bibfield  {journal} {\bibinfo  {journal} {Class. Quant. Grav.}\ }\textbf
  {\bibinfo {volume} {40}},\ \bibinfo {pages} {115006} (\bibinfo {year}
  {2023})},\ \Eprint {https://arxiv.org/abs/2212.02559} {arXiv:2212.02559
  [hep-th]} \BibitemShut {NoStop}%
\bibitem [{\citenamefont {Gao}\ and\ \citenamefont {Wald}(2000)}]{Gao:2000ga}%
  \BibitemOpen
  \bibfield  {author} {\bibinfo {author} {\bibfnamefont {S.}~\bibnamefont
  {Gao}}\ and\ \bibinfo {author} {\bibfnamefont {R.~M.}\ \bibnamefont {Wald}},\
  }\bibfield  {title} {\bibinfo {title} {{Theorems on gravitational time delay
  and related issues}},\ }\href {https://doi.org/10.1088/0264-9381/17/24/305}
  {\bibfield  {journal} {\bibinfo  {journal} {Class. Quant. Grav.}\ }\textbf
  {\bibinfo {volume} {17}},\ \bibinfo {pages} {4999} (\bibinfo {year}
  {2000})},\ \Eprint {https://arxiv.org/abs/gr-qc/0007021}
  {arXiv:gr-qc/0007021} \BibitemShut {NoStop}%
\bibitem [{\citenamefont {Aichelburg}\ and\ \citenamefont
  {Sexl}(1971)}]{Aichelburg:1970dh}%
  \BibitemOpen
  \bibfield  {author} {\bibinfo {author} {\bibfnamefont {P.~C.}\ \bibnamefont
  {Aichelburg}}\ and\ \bibinfo {author} {\bibfnamefont {R.~U.}\ \bibnamefont
  {Sexl}},\ }\bibfield  {title} {\bibinfo {title} {{On the Gravitational field
  of a massless particle}},\ }\href {https://doi.org/10.1007/BF00758149}
  {\bibfield  {journal} {\bibinfo  {journal} {Gen. Rel. Grav.}\ }\textbf
  {\bibinfo {volume} {2}},\ \bibinfo {pages} {303} (\bibinfo {year}
  {1971})}\BibitemShut {NoStop}%
\bibitem [{\citenamefont {Hotta}\ and\ \citenamefont
  {Tanaka}(1993)}]{Hotta:1992qy}%
  \BibitemOpen
  \bibfield  {author} {\bibinfo {author} {\bibfnamefont {M.}~\bibnamefont
  {Hotta}}\ and\ \bibinfo {author} {\bibfnamefont {M.}~\bibnamefont {Tanaka}},\
  }\bibfield  {title} {\bibinfo {title} {{Shock wave geometry with nonvanishing
  cosmological constant}},\ }\href {https://doi.org/10.1088/0264-9381/10/2/012}
  {\bibfield  {journal} {\bibinfo  {journal} {Class. Quant. Grav.}\ }\textbf
  {\bibinfo {volume} {10}},\ \bibinfo {pages} {307} (\bibinfo {year}
  {1993})}\BibitemShut {NoStop}%
\bibitem [{\citenamefont {Horowitz}\ and\ \citenamefont
  {Itzhaki}(1999)}]{Horowitz:1999gf}%
  \BibitemOpen
  \bibfield  {author} {\bibinfo {author} {\bibfnamefont {G.~T.}\ \bibnamefont
  {Horowitz}}\ and\ \bibinfo {author} {\bibfnamefont {N.}~\bibnamefont
  {Itzhaki}},\ }\bibfield  {title} {\bibinfo {title} {{Black holes, shock
  waves, and causality in the AdS / CFT correspondence}},\ }\href
  {https://doi.org/10.1088/1126-6708/1999/02/010} {\bibfield  {journal}
  {\bibinfo  {journal} {JHEP}\ }\textbf {\bibinfo {volume} {02}},\ \bibinfo
  {pages} {010}},\ \Eprint {https://arxiv.org/abs/hep-th/9901012}
  {arXiv:hep-th/9901012} \BibitemShut {NoStop}%
\bibitem [{\citenamefont {Gubser}\ \emph {et~al.}(2008)\citenamefont {Gubser},
  \citenamefont {Pufu},\ and\ \citenamefont {Yarom}}]{Gubser:2008pc}%
  \BibitemOpen
  \bibfield  {author} {\bibinfo {author} {\bibfnamefont {S.~S.}\ \bibnamefont
  {Gubser}}, \bibinfo {author} {\bibfnamefont {S.~S.}\ \bibnamefont {Pufu}},\
  and\ \bibinfo {author} {\bibfnamefont {A.}~\bibnamefont {Yarom}},\ }\bibfield
   {title} {\bibinfo {title} {{Entropy production in collisions of
  gravitational shock waves and of heavy ions}},\ }\href
  {https://doi.org/10.1103/PhysRevD.78.066014} {\bibfield  {journal} {\bibinfo
  {journal} {Phys. Rev. D}\ }\textbf {\bibinfo {volume} {78}},\ \bibinfo
  {pages} {066014} (\bibinfo {year} {2008})},\ \Eprint
  {https://arxiv.org/abs/0805.1551} {arXiv:0805.1551 [hep-th]} \BibitemShut
  {NoStop}%
\bibitem [{\citenamefont {Caron-Huot}\ \emph {et~al.}(2021)\citenamefont
  {Caron-Huot}, \citenamefont {Mazac}, \citenamefont {Rastelli},\ and\
  \citenamefont {Simmons-Duffin}}]{Caron-Huot:2021rmr}%
  \BibitemOpen
  \bibfield  {author} {\bibinfo {author} {\bibfnamefont {S.}~\bibnamefont
  {Caron-Huot}}, \bibinfo {author} {\bibfnamefont {D.}~\bibnamefont {Mazac}},
  \bibinfo {author} {\bibfnamefont {L.}~\bibnamefont {Rastelli}},\ and\
  \bibinfo {author} {\bibfnamefont {D.}~\bibnamefont {Simmons-Duffin}},\
  }\bibfield  {title} {\bibinfo {title} {{Sharp boundaries for the
  swampland}},\ }\href {https://doi.org/10.1007/JHEP07(2021)110} {\bibfield
  {journal} {\bibinfo  {journal} {JHEP}\ }\textbf {\bibinfo {volume} {07}},\
  \bibinfo {pages} {110}},\ \Eprint {https://arxiv.org/abs/2102.08951}
  {arXiv:2102.08951 [hep-th]} \BibitemShut {NoStop}%
\bibitem [{\citenamefont {Kabat}\ and\ \citenamefont
  {Ortiz}(1992)}]{Kabat:1992tb}%
  \BibitemOpen
  \bibfield  {author} {\bibinfo {author} {\bibfnamefont {D.~N.}\ \bibnamefont
  {Kabat}}\ and\ \bibinfo {author} {\bibfnamefont {M.}~\bibnamefont {Ortiz}},\
  }\bibfield  {title} {\bibinfo {title} {{Eikonal quantum gravity and Planckian
  scattering}},\ }\href {https://doi.org/10.1016/0550-3213(92)90627-N}
  {\bibfield  {journal} {\bibinfo  {journal} {Nucl. Phys. B}\ }\textbf
  {\bibinfo {volume} {388}},\ \bibinfo {pages} {570} (\bibinfo {year}
  {1992})},\ \Eprint {https://arxiv.org/abs/hep-th/9203082}
  {arXiv:hep-th/9203082} \BibitemShut {NoStop}%
\bibitem [{\citenamefont {Kologlu}\ \emph {et~al.}(2020)\citenamefont
  {Kologlu}, \citenamefont {Kravchuk}, \citenamefont {Simmons-Duffin},\ and\
  \citenamefont {Zhiboedov}}]{Kologlu:2019bco}%
  \BibitemOpen
  \bibfield  {author} {\bibinfo {author} {\bibfnamefont {M.}~\bibnamefont
  {Kologlu}}, \bibinfo {author} {\bibfnamefont {P.}~\bibnamefont {Kravchuk}},
  \bibinfo {author} {\bibfnamefont {D.}~\bibnamefont {Simmons-Duffin}},\ and\
  \bibinfo {author} {\bibfnamefont {A.}~\bibnamefont {Zhiboedov}},\ }\bibfield
  {title} {\bibinfo {title} {{Shocks, Superconvergence, and a Stringy
  Equivalence Principle}},\ }\href {https://doi.org/10.1007/JHEP11(2020)096}
  {\bibfield  {journal} {\bibinfo  {journal} {JHEP}\ }\textbf {\bibinfo
  {volume} {11}},\ \bibinfo {pages} {096}},\ \Eprint
  {https://arxiv.org/abs/1904.05905} {arXiv:1904.05905 [hep-th]} \BibitemShut
  {NoStop}%
\end{thebibliography}%

\end{document}